\documentstyle[twoside,fleqn,espcrc2,amssymb,epsfig]{article}
\title{Neutrino masses and mixing
from neutrino oscillation data}
\author{
S.M. Bilenky\address{Joint Institute for Nuclear Research, Dubna, Russia},
C. Giunti\address{INFN, Sezione di Torino, 
and
Dipartimento di Fisica Teorica,
Via P. Giuria 1, I--10125 Torino, Italy}
and
W. Grimus\address{Institute for Theoretical Physics, University of Vienna,
Boltzmanngasse 5, A--1090 Vienna, Austria}
}
\begin{document}
\begin{abstract}
We discuss what information
about the neutrino mass spectrum and the elements
of the neutrino mixing matrix
can be inferred from the results of neutrino oscillation experiments.
The cases of three and four massive neutrinos are considered.
It is shown that the general characteristics of the neutrino
mixing matrix are quite different from those of the
quark mixing matrix.
Talk presented by S.M. Bilenky at the
XVI \emph{International Workshop on Weak Interactions
and Neutrinos},
Capri, Italy, June 22--28 1997.
Preprint DFTT 45/97, hep-ph/9707358.
\end{abstract}
\maketitle

The problem of neutrino masses and mixing
is the central issue of today's neutrino physics.
The results of many experiments
on the search for effects of neutrino masses and mixing
were discussed at this meeting.
New and more precise experiments
are going on or are under preparation.

We will present here some results
of Refs.\cite{BBGK,BGKP,BGG96,BGG97},
in which
model independent information about the spectrum of neutrino masses and
the elements of the neutrino mixing matrix
was obtained from the existing 
neutrino oscillation data.
We will consider here also possible implications 
of our results for the future experiments.

In accordance with the neutrino mixing hypothesis
\cite{Pontecorvo57},
the left-handed flavor neutrino fields
$\nu_{{\alpha}L}$
determined by the standard CC and NC interactions
are mixtures of the left-handed components
of the fields $\nu_i$ of neutrinos (Dirac or Majorana) with 
mass $m_i$:
\begin{equation}
\nu_{{\alpha}L}
=
\sum_{i=1}^{n}
U_{{\alpha}i} \, \nu_{iL}
\,,
\label{01}
\end{equation}
where $U^{\dagger}U=1$,
$\alpha=e,\mu,\tau,\ldots$ and $n\geq3$.

From the existing neutrino oscillation data
it follows that three different scales
of neutrino mass-squared differences can be relevant for
neutrino oscillations:
$\simeq10^{-5}\,\mbox{eV}^2$ for solar neutrinos
\cite{solar},
$\simeq10^{-2}\,\mbox{eV}^2$ for atmospheric neutrinos
\cite{atmospheric}
and
$\simeq1\,\mbox{eV}^2$ for LSND neutrinos
\cite{LSND}.
Let us consider the general case of $n$ neutrinos
($m_1<m_2<\ldots<m_n$)
with the largest mass-squared difference
$ \Delta{m}^{2} \equiv m_n^2 - m_1^2 $
relevant for short-baseline (SBL) oscillations
and two groups of close masses
$m_1<\ldots<m_{r-1}$
and
$m_r<\ldots<m_n$
such that
\begin{eqnarray}
&&
\Delta{m}^{2}_{i1}
\ll
\Delta{m}^{2}
\quad \mbox{for} \quad
i < r
\,,
\nonumber
\\
&&
\Delta{m}^{2}_{nj}
\ll
\Delta{m}^{2}
\quad \mbox{for} \quad
j \geq r
\,,
\label{02}
\end{eqnarray}
with
$ \Delta{m}^{2}_{ij} \equiv m_i^2 - m_j^2 $.
In this case,
for the SBL transition probabilities
we have \cite{BGKP}
\begin{eqnarray}
&&
P^{({\rm SBL})}_{\stackrel{\makebox[0pt][l]
{$\hskip-3pt\scriptscriptstyle(-)$}}{\nu_{\alpha}}
\to\stackrel{\makebox[0pt][l]
{$\hskip-3pt\scriptscriptstyle(-)$}}{\nu_{\beta}}}
=
\frac{A_{\alpha;\beta}}{2}
\left( 1 - \cos \frac{ \Delta{m}^{2} L }{ 2 p } \right)
\quad
(\beta\neq\alpha)
,
\label{03}
\\
&&
P^{({\rm SBL})}_{\stackrel{\makebox[0pt][l]
{$\hskip-3pt\scriptscriptstyle(-)$}}{\nu_{\alpha}}
\to\stackrel{\makebox[0pt][l]
{$\hskip-3pt\scriptscriptstyle(-)$}}{\nu_{\alpha}}}
=
1 - \frac{B_{\alpha;\alpha}}{2}
\left(1 - \cos \frac{ \Delta{m}^{2} L }{ 2 p } \right)
,
\label{04}
\end{eqnarray}
where $p$ is the neutrino momentum and $L$ is the distance between
the neutrino source and detector.
The oscillation amplitudes are given by
\begin{eqnarray}
&&
A_{\alpha;\beta}
=
4 \left| \sum_{i=r}^{n} U_{{\beta}i} \, U_{{\alpha}i}^{*} \right|^2
,
\label{05}
\\
&&
B_{\alpha;\alpha}
=
4
\left( \sum_{i=r}^{n} |U_{{\alpha}i}|^2 \right)
\left( 1 - \sum_{i=r}^{n} |U_{{\alpha}i}|^2 \right)
.
\label{06}
\end{eqnarray}

No indication in favour of neutrino oscillations
was found
in SBL reactor and accelerator disappearance experiments.
From the exclusion plots obtained from the results of these
experiments it follows that
\begin{equation}
B_{e;e} \leq B_{e;e}^{0}
\quad \mbox{and} \quad
B_{\mu;\mu} \leq B_{\mu;\mu}^{0}
\,.
\label{07}
\end{equation}
The numerical values of the upper bounds
$B_{\alpha;\alpha}^{0}$
($\alpha=e,\mu$)
depend on the value of
$\Delta{m}^2$.
We considered the wide interval
\begin{equation}
10^{-1}
\leq
\Delta{m}^2
\leq
10^{3} \, \mbox{eV}^2
\,.
\label{08}
\end{equation}
From Eqs.(\ref{06}) and (\ref{07}) it follows that
\begin{equation}
\sum_{i=r}^{n}
|U_{{\alpha}i}|^2
\leq
a^0_\alpha
\quad \mbox{or} \quad
\sum_{i=r}^{n}
|U_{{\alpha}i}|^2
\geq
1 - a^0_\alpha
\,,
\label{09}
\end{equation}
with
$\alpha=e,\mu$
and
\begin{equation}
a^0_\alpha
=
\frac{1}{2}
\left(
1
-
\sqrt{ 1 - B_{\alpha;\alpha}^{0} }
\right)
\,.
\label{10}
\end{equation}
From the results of
the Bugey
$\bar\nu_e\to\bar\nu_e$
\cite{Bugey95}
and
the
CDHS and CCFR
\cite{CDHS84-CCFR84}
$
\stackrel{\makebox[0pt][l]
{$\hskip-3pt\scriptscriptstyle(-)$}}{\nu_{\mu}}
\to
\stackrel{\makebox[0pt][l]
{$\hskip-3pt\scriptscriptstyle(-)$}}{\nu_{\mu}}
$
experiments it follows that the values of
$a^0_e$ and $a^0_\mu$
are small
($ a^{0}_e \lesssim 4 \times 10^{-2} $
for any value of
$\Delta{m}^{2}$
in the interval (\ref{08})
and
$ a^{0}_\mu \lesssim 2 \times 10^{-1} $
for
$
\Delta{m}^{2} \gtrsim 0.3 \, \mbox{eV}^2
$).
Thus, from the existing inclusive data it 
follows that
the sums
$
\sum_{i=r}^{n}
|U_{ei}|^2
$
and
$
\sum_{i=r}^{n}
|U_{{\mu}i}|^2
$
can be either small or large (i.e., close to one).

\begin{figure}[t]
\begin{center}
\mbox{\epsfig{file=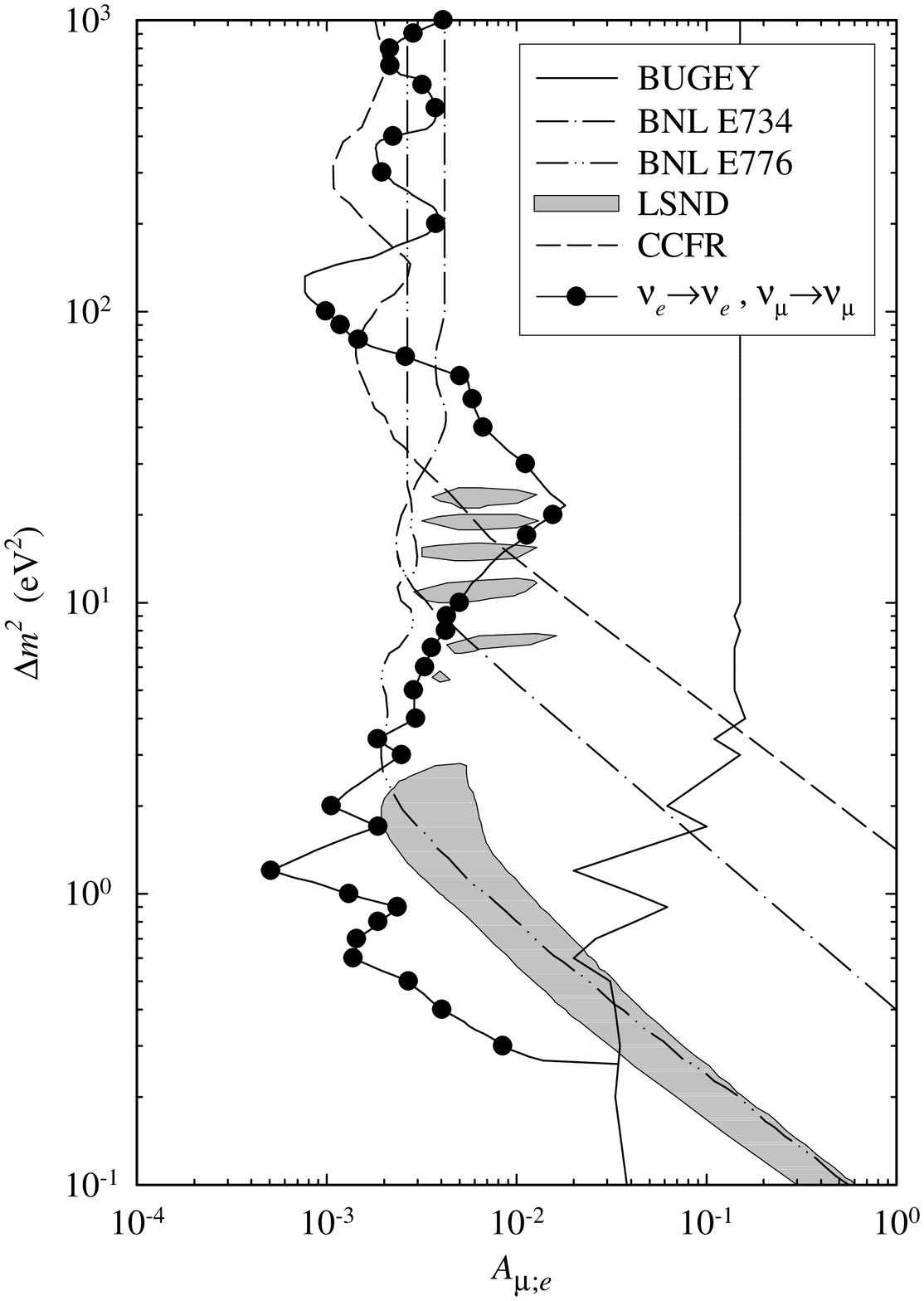,width=\linewidth}}
\\
Figure \ref{fig1}
\end{center}
\null
\vspace{-1.5cm}
\null
\refstepcounter{figure}
\label{fig1}
\end{figure}

We will consider first the simplest case of three massive neutrinos
with the mass hierarchy
$ m_1 \ll m_2 \ll m_3 $
and will assume that
$ \Delta{m}^2_{21} $
is relevant for the
suppression of solar $\nu_e$'s.
In this case, we have
\begin{equation}
|U_{{\alpha}3}|^2
\leq
a^0_\alpha
\quad \mbox{or} \quad
|U_{{\alpha}3}|^2
\geq
1 - a^0_\alpha
\,,
\label{11}
\end{equation}
with $\alpha=e,\mu$.
In the case under consideration,
the probability of solar $\nu_e$'s to survive
is given by \cite{SS92}
\begin{equation}
P^{{\rm sun}}_{\nu_e\to\nu_e}
=
|U_{e3}|^4
+
\left( 1 - |U_{e3}|^2 \right)^2
P^{(1;2)}_{\nu_e\to\nu_e}
\,,
\label{12}
\end{equation}
where $P^{(1;2)}_{\nu_e\to\nu_e}$ is the probability of $\nu_e$'s
to survive
due to the coupling of $\nu_e$ with $\nu_1$ and $\nu_2$.
From Eq.(\ref{12}) it follows that,
in order to explain the solar neutrino data \cite{solar}, 
from the two possibilities (\ref{11}) for
$|U_{e3}|^2$
we must
choose
$ |U_{e3}|^2 \leq a^0_e $.
Hence, we come to the two possible schemes:
\begin{eqnarray}
&&
(\mbox{I})
\quad
|U_{e3}|^2 \leq a^0_e
\,,
\quad
|U_{\mu3}|^2 \leq a^0_\mu
\,;
\label{13}
\\
&&
(\mbox{II})
\quad
|U_{e3}|^2 \leq a^0_e
\,,
\quad
|U_{\mu3}|^2 \geq 1 - a^0_\mu
\,.
\label{14}
\end{eqnarray}
The most natural scheme I,
with the 
hierarchy of couplings
corresponding to the hierarchy of neutrino masses,
is not favoured by the results of the LSND experiment.
In fact,
from
Eqs.(\ref{05}) and (\ref{13}) we have the following upper bound
for the amplitude
$
A_{\mu;e}
=
4 |U_{e3}|^2 |U_{\mu3}|^2
$
of
$
\stackrel{\makebox[0pt][l]
{$\hskip-3pt\scriptscriptstyle(-)$}}{\nu_{\mu}}
\to
\stackrel{\makebox[0pt][l]
{$\hskip-3pt\scriptscriptstyle(-)$}}{\nu_{e}}
$
transitions:
\begin{equation}
A_{\mu;e}
\leq
4 a^0_e a^0_\mu
\,.
\label{15}
\end{equation}
As it is seen from Fig.\ref{fig1},
the LSND-allowed region that is not excluded
by the data of other experiments
lies inside of the region forbidden by the inequality
(\ref{15})
(the curve passing through the circles).

On the other hand,
scheme II allows to describe the results of all experiments,
including LSND.
Note that if this scheme is realized in nature we have
$
m_{\nu_\mu} \gg m_{\nu_e} , m_{\nu_\tau}
$.

Up to now we did not take into account
the atmospheric neutrino anomaly
\cite{atmospheric}.
In order to take into account all data
we will consider schemes with four massive neutrinos
\cite{BGKP,BGG96,BGG97}.
Let us first assume that there is a hierarchy of neutrino masses,
$ m_1 \ll m_2 \ll m_3 \ll m_4 $,
with
$\Delta{m}^{2}_{21}$
relevant for the suppression of solar $\nu_e$'s,
$\Delta{m}^{2}_{31}$
relevant for the atmospheric neutrino anomaly and
$\Delta{m}^{2}\equiv\Delta{m}^{2}_{41}$
relevant for the LSND anomaly.
From the data of inclusive experiments
it follows that
\begin{equation}
|U_{{\alpha}4}|^2
\leq
a^0_\alpha
\quad \mbox{or} \quad
|U_{{\alpha}4}|^2
\geq
1 - a^0_\alpha
\,,
\label{16}
\end{equation}
with $\alpha=e,\mu$.
The survival probability of solar $\nu_e$'s
is given by Eq.(\ref{12})
with the obvious change
$ |U_{e3}| \to |U_{e4}| $.
For the survival probability of atmospheric
$
\stackrel{\makebox[0pt][l]
{$\hskip-3pt\scriptscriptstyle(-)$}}{\nu_{\mu}}
$'s
we have the lower bound
\begin{equation}
P_{\stackrel{\makebox[0pt][l]
{$\hskip-3pt\scriptscriptstyle(-)$}}{\nu_{\mu}}
\to
\stackrel{\makebox[0pt][l]
{$\hskip-3pt\scriptscriptstyle(-)$}}{\nu_{\mu}}}
\geq
|U_{\mu4}|^2
\,.
\label{17}
\end{equation}
Thus, from the solar and atmospheric neutrino data it follows that
both elements
$|U_{e4}|^2$
and
$|U_{\mu4}|^2$
are small:
\begin{equation}
|U_{e4}|^2
\leq
a^0_e
\quad \mbox{and} \quad
|U_{\mu4}|^2
\leq
a^0_\mu
\,.
\label{18}
\end{equation}
Hence,
for the amplitude
$
A_{\mu;e}
=
4 |U_{e4}|^2 |U_{\mu4}|^2
$
of
$
\stackrel{\makebox[0pt][l]
{$\hskip-3pt\scriptscriptstyle(-)$}}{\nu_{\mu}}
\to
\stackrel{\makebox[0pt][l]
{$\hskip-3pt\scriptscriptstyle(-)$}}{\nu_{e}}
$
transitions we have the upper bound (\ref{15})
that is presented in
Fig.\ref{fig1}
(the curve passing through the circles).
From the figure one can see that
this limit is in contradiction with
the LSND-allowed region.
Thus, we come to the conclusion that the hierarchy of masses
of four neutrinos is not favoured by the results
of the LSND and other neutrino
oscillation experiments.
The same conclusion is valid for all
the possible neutrino mass spectra with one mass separated from 
the group of three close masses by a gap
of about 1 eV.

Only the following two mass spectrum of four
neutrinos are favoured by all the existing 
neutrino oscillation data:
\begin{eqnarray}
&&
\mbox{(A)}
\qquad
\underbrace{
\overbrace{m_1 < m_2}^{{\rm atm}}
\ll
\overbrace{m_3 < m_4}^{{\rm solar}}
}_{{\rm LSND}}
\,,
\label{20}
\\
&&
\mbox{(B)}
\qquad
\underbrace{
\overbrace{m_1 < m_2}^{{\rm solar}}
\ll
\overbrace{m_3 < m_4}^{{\rm atm}}
}_{{\rm LSND}}
\,.
\label{21}
\end{eqnarray}
Taking into account the data of the solar and atmospheric neutrino
experiments, for the schemes A and B we have respectively
\begin{eqnarray}
&&
\mbox{(A)}
\quad
\sum_{i=1,2}
|U_{ei}|^2 \leq a^0_e
\,,
\quad
\sum_{i=3,4}
|U_{{\mu}i}|^2 \leq a^0_\mu
\,,
\label{22}
\\
&&
\mbox{(B)}
\quad
\sum_{i=3,4}
|U_{ei}|^2 \leq a^0_e
\,,
\quad
\sum_{i=1,2}
|U_{{\mu}i}|^2 \leq a^0_\mu
\,.
\label{23}
\end{eqnarray}
For the amplitude of 
$
\stackrel{\makebox[0pt][l]
{$\hskip-3pt\scriptscriptstyle(-)$}}{\nu_{\mu}}
\to
\stackrel{\makebox[0pt][l]
{$\hskip-3pt\scriptscriptstyle(-)$}}{\nu_{e}}
$
oscillations
in both schemes we have the upper bound 
\begin{equation}
A_{\mu;e}
\leq
4 \, \mbox{Min}(a^0_e,a^0_\mu)
\,,
\label{24}
\end{equation}
which is compatible with the LSND result.

If scheme A is realized in nature,
the experiments on the measurement of the neutrino mass
with the $^3$H method and
the experiments on the search for
neutrinoless double-beta decay have a good chance
to see the effects of the SBL mass-squared difference
$\Delta{m}^{2}$.
In fact, taking into account the inequalities
(\ref{22}),
for the mass $m(^3\mbox{H})$
and for the effective Majorana mass
$
\langle{m}\rangle
=
\sum_{i=1}^{n} U_{ei}^2 m_i
$
in scheme A we have
\begin{eqnarray}
&&
m(^3\mbox{H})
\simeq
\sqrt{ \Delta{m}^{2} }
\,,
\label{25}
\\
&&
|\langle{m}\rangle|
\simeq
\left| \sum_{i=3,4} U_{ei}^2 \right|
\sqrt{ \Delta{m}^{2} }
\leq
\sqrt{ \Delta{m}^{2} }
\,.
\label{26}
\end{eqnarray}

In the framework of the schemes A and B,
rather strong constraints for the probabilities of long-baseline (LBL)
$\bar\nu_e\to\bar\nu_e$
and
$
\stackrel{\makebox[0pt][l]
{$\hskip-3pt\scriptscriptstyle(-)$}}{\nu_{\mu}}
\to\stackrel{\makebox[0pt][l]
{$\hskip-3pt\scriptscriptstyle(-)$}}{\nu_{e}}
$
transitions
can be inferred from the bounds on the
elements of the mixing matrix
obtained from SBL experiments
\cite{BGG97}.
Let us consider scheme A.
For
the probabilities of
$\nu_{\alpha}\to\nu_{\beta}$
transitions in LBL experiments
we have
\begin{eqnarray}
P^{({\rm LBL,A})}_{\nu_\alpha\to\nu_\beta}
=
\null
&&
\null
\left|
\sum_{k=1,2}
U_{{\beta}k}
\,
U_{{\alpha}k}^{*}
\,
e^{ - i \frac{ \Delta{m}^{2}_{k1} L }{ 2 p } }
\right|^2
\nonumber
\\
&&
\null
+
\left|
\sum_{k=3,4}
U_{{\beta}k}
\,
U_{{\alpha}k}^{*}
\right|^2
\,.
\label{29}
\end{eqnarray}
The corresponding equation for antineutrinos
implies that
the probability of LBL
$\bar\nu_e\to\bar\nu_e$
transitions is bounded by
\begin{equation}
P^{({\rm LBL})}_{\bar\nu_e\to\bar\nu_e}
\geq
\left( 1 - \sum_{i=1,2} |U_{ei}|^2 \right)^2
\,.
\label{31}
\end{equation}
From the results of SBL reactor experiments
it follows that
$
\sum_{i=1,2} |U_{ei}|^2
$
is small (see Eq.(\ref{22}))
and thus (in the scheme under consideration)
the LBL probability
$
P^{({\rm LBL})}_{\bar\nu_e\to\bar\nu_e}
$
must be close to one.
In fact,
from Eqs.(\ref{22}) and (\ref{31}),
for the probability of the LBL
transitions of $\bar\nu_e$'s into all possible states
we have the upper bound
\begin{equation}
1 - P^{({\rm LBL})}_{\bar\nu_e\to\bar\nu_e}
\leq
a^0_e \left( 2 - a^0_e \right)
\,.
\label{32}
\end{equation}
The same inequality is valid in scheme B.

\begin{figure}[t]
\begin{center}
\mbox{\epsfig{file=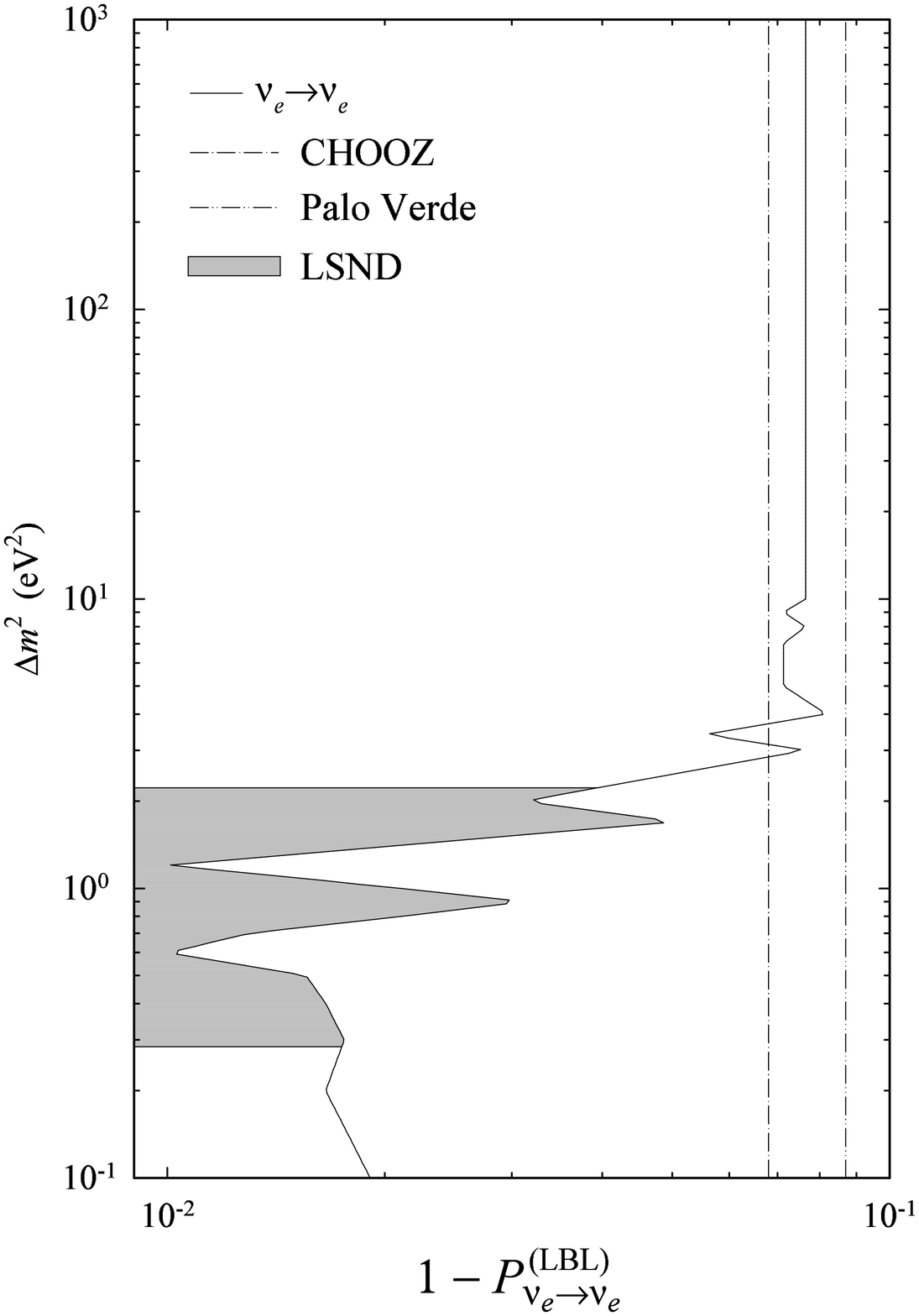,width=\linewidth}}
\\
Figure \ref{fig2}
\end{center}
\null
\vspace{-1.5cm}
\null
\refstepcounter{figure}
\label{fig2}
\end{figure}

The upper bound (\ref{32}) is presented in Fig.\ref{fig2}
for the values of the SBL parameter
$\Delta{m}^{2}$
in the range (\ref{08}) (the solid line).
The dash-dotted
and dash-dot-dotted lines correspond to
the expected sensitivity
of the CHOOZ and Palo Verde LBL reactor experiments
\cite{CHOOZ-PaloVerde}.
As it is
seen from Fig.\ref{fig2},
the sensitivity of the
CHOOZ experiment
could allow to see effect of neutrino oscillations only if
$ \Delta{m}^{2} \gtrsim 3 \, \mbox{eV}^2 $.
The shadowed region in Fig.\ref{fig2}
corresponds to the range of $\Delta{m}^2$
allowed by the results of the LSND experiment,
taking into account the results of
all the other SBL experiments.

In conclusion, we have considered
here the following question: what information
about the neutrino mass spectrum and the elements
of the neutrino mixing matrix can be inferred from the results
of neutrino oscillation experiments? We came to
the conclusion that from the existing data it follows that
the general characteristics of neutrino and
quark mixing are quite different.
Only
future neutrino oscillation experiments will
allow to obtain information about the genuine neutrino
mass spectrum and the neutrino mixing matrix.

\end{document}